# The Drawing You Watch Is the Portrait You Keep: A Replayable Machine Portrait Studio

**Vinod Rajpurohit**

rajp0001@e.ntu.edu.sg

ABSTRACT

Ask an image model for your portrait and a finished picture arrives: no first line, no second thoughts, no record of how it got there, and a different picture if you ask again. I built a portrait studio that works the other way. A sitter chooses one of eight fictional artists, each written with a city, a medium and a tempo of its own, holds still for 8 to 12 seconds, and watches a drawing of themselves appear over 44 seconds. One still is sent after consent, a hosted image model returns a private study the sitter never sees, and a compiler turns that study into one drawing document. The live drawing, the saved sheet, the vector export and the film all read that document, and a replay makes no model call, so the drawing you watch is the portrait you keep. I describe the compiler in enough detail to rebuild it and say plainly what it is: an authored reconstruction at a compressed speed, not a recorded hand. On 16 generated test studies a sheet holds 8,784 to 16,378 timed contacts, about three quarters of them fixed-grid fills and nearly all shorter than one video frame, and recompiling one study gave byte-identical output four times on one machine. The paper also traces a bug that matters more than any of that. The earlier compiler estimated the paper colour tile by tile, read dark skin as paper, and in all eight hands kept less of a deep-complexioned face than a fair one (13 to 48 percent against 41 to 56). On 80 generated studies the rebuilt engine keeps the complexion difference the study already has (raw retention 89 to 120 percent), so the fairness question has moved upstream, to the study and the hosted model. The two test sitters are generated and differ in more than complexion, and no real-face or perceptual study exists yet. The studio is live at https://everyfaceproject.com.

## 1. Introduction

I started from one observation: a machine can draw now. When I paid attention to how it draws, two things bothered me. It draws in one shot. The picture arrives finished, with no first line, no second thoughts and no record of how it got there. And it cannot do it twice. Ask for the same drawing again and a different drawing comes back. A drawing by a person is nothing but history: where the pencil started, what got corrected, the order things happened in, the signature at the end. So the project became a machine drawing with a history, one you can watch being made and replay exactly, as many times as you like. Once the drawing had a history, the person being drawn needed one too, so I designed a sitting around it (Section 2).

There is also the person being drawn. Sitting for a portrait is an odd, exposed few seconds: you hold still, you are looked at, and you do not get to choose what the person drawing you sees. That exposure is what makes a portrait worth keeping, and it is exactly what a regenerate button takes away. A picture you can redo forever stops being a picture of you. So the studio has no prompt box, no style slider and no retake.

The eight artists are the other half of the design, and they were not picked at random. Each one is written down before any sitter arrives: a city, a medium that belongs to a real drawing tradition, a number of seconds it looks before it chooses, a short list of what it notices first and what it leaves alone, and a few lines it says while it looks. They came from watching people draw, from conversations with artists about what they look at before the first mark, and from how marks actually go down in each tradition: a sanguine crayon lands once and stays, a Jaipur kalam line is laid faint and then committed, a Nanyang brush thickens where the form turns. The tempo is the personality. Sol in New York looks for 8 seconds, because that is how a reportage hand works; Yeon in Seoul looks for 12, because the Joseon rule she quotes says that one hair wrong is another person. Section 2 gives them in full, and Figure 2 shows all eight drawing the same sitter.

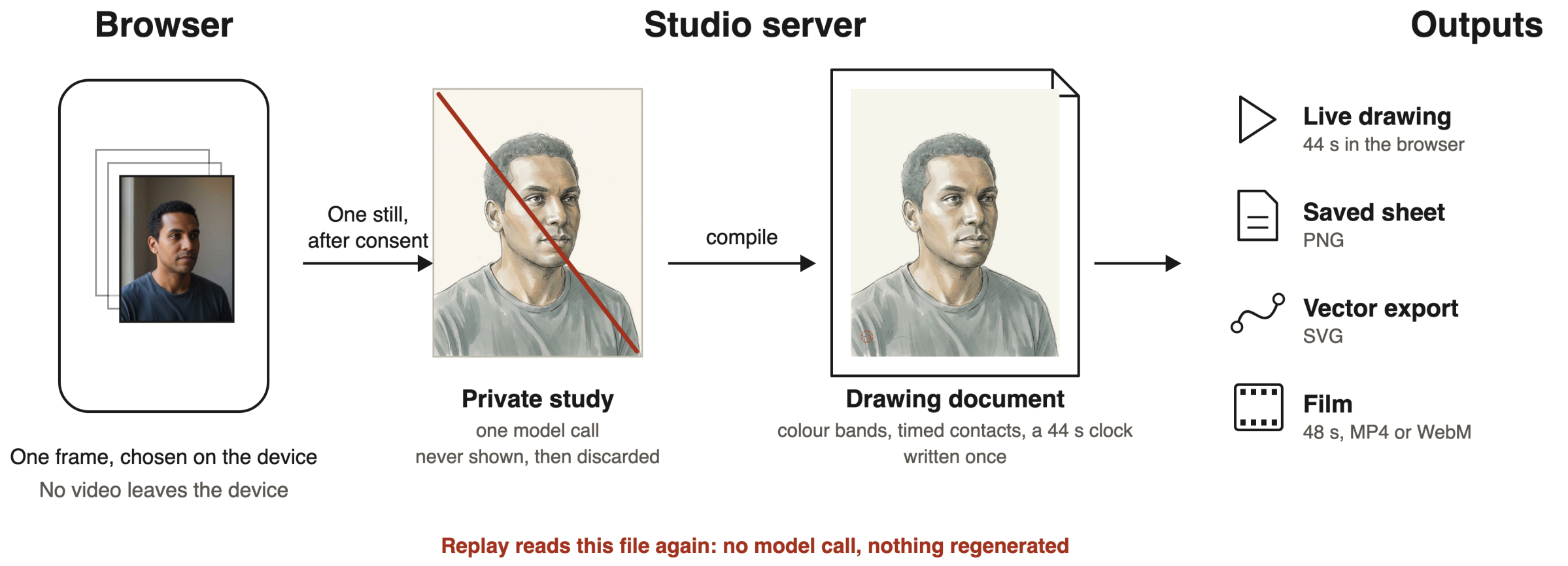

**Figure 1:** The pipeline, shown with the studio's fictional demonstration sitter. In the browser, camera frames are scored on the device and one still is sent, once per attempt, after consent. On the server, one call to a hosted image model returns a private study (here Mira's committed reference study, struck through because a sitter never sees it and it is then discarded), and the compiler writes one drawing document (here Mira's finished sheet). Four surfaces read that document: the live drawing, the saved sheet, the vector export and the film. Replay makes no model call.

That idea has a limit. Replay is exact because the drawing document is fixed once it is written. The hosted image model that makes the study is not seeded and can return a different study each time it is asked, so the history begins at the study, not at the sitter.

A finished image can be animated with a wipe, a fade or a scribble timed to look plausible, and nothing ties that animation to the picture that is kept. The constraint I set for this studio is that the drawing the sitter watches and the portrait they keep are one document, read by the live player, the saved sheet, the vector export and the film, with nothing regenerated for a replay. The current engine has some of that same structure: its colour regions are computed before the first contact plays, and the contacts reveal them. The difference is that the document revealing them is the one the sitter keeps, and the product calls it a reconstruction: a performance authored by rule from a finished study, not recovered from any real process. The order is set by fixed rules over the study's own marks: the traced ink and the borrowed paths follow them, and the grid returns, about three quarters of the contacts, follow a raster scan.

The compiler has had two generations, and this paper describes both because the complexion history runs through both. The earlier compiler measured darkness in the study against a locally estimated paper level and rebuilt it as lines and hatching in the artist's own ink; its sheets were lighter and less faithful than the studies they came from. For a sitter with darker skin, that meant a face drawn fainter than their own study, with no error and no warning. The current engine keeps the study's own pigment. It quantises the non-paper pixels to a small palette, traces the boundary of every colour band as a polygon, and schedules eight to sixteen thousand timed contacts that reveal those bands in an order set by rules. The sheet is now the study, quantised, with a signature, and the performance is an authored reconstruction of how such a sheet might be drawn. Fidelity and colour now come from the study, so the compiler can no longer narrow or correct what the study got wrong, complexion included.

This paper reports four things. First, the engine as it runs in the live studio: selected by an environment flag, unchanged in source at the commit this paper describes, and with its contacts stage specified in enough detail to reimplement. That specification covers the paper estimate, the ink and tone split, the per-material palette, the vector boundaries, the ink tracing, the borrowed tone directions, the grid fill, the pass routing, the timing formula and the budgets. Three of those inputs still depend on the earlier compiler, which Section 4.3 only outlines and Section 4.2 names. The parts that are classical are named as classical: thinning, distance transforms, k-means colour quantisation and pixel-boundary vectorisation are decades old. Second, measurements of what the document contains and how far its determinism has been checked, with the scope of each measurement stated. Third, a complexion trace across the two compilers: the mechanism by which an adaptive local background estimate keeps less of a dark face, the fixes, and a same-day re-measurement on 80 generated studies. That re-measurement shows the current engine preserving the study's own complexion separation. Fourth, the practice of pinning the product's claims about itself with tests, reported together with the places it failed.

## 2. The sitting

From the sitter's side it goes like this. You pick an artist from eight. Each is fictional, has a city, a medium and a way of working, and speaks a few written lines while observing you. Those lines are authored ahead of time; none of them is a reading of your face. The observation lasts 8 to 12 seconds depending on the artist: Sol in New York looks for 8, Kavi in Chennai and Halia in Singapore for 9, Odile in Paris and Aya in Tokyo for 10, Mira in San Francisco and Ganga in Jaipur for 11, and Yeon in Seoul for 12. During that window your browser samples frames and scores them for image quality on your device. The live video is not uploaded. One still is sent, after you have consented, once per attempt.

Then there is a short wait while the study is made and compiled, separate from the drawing: the server's preparation deadline is 75 seconds, and your browser gives up at 95. Then the drawing plays over 44 seconds in seven passes, each with a name you can read as it runs, and the sheet is signed with the artist's initial and the sitting's edition number. You never see the study; the sheet you keep is rendered from the drawing document. You can watch it again at half, normal or double speed, download it as a vector file, and record a 48 second film of it being drawn. There is no retake for preference. A sitting gets at most two generation attempts in all, and a refusal the studio owns is not counted against them; a finished portrait is not redrawn because you would have liked another expression. If it catches you tired, you keep the portrait of you tired. The app does not keep the still.

### *2.1 The eight artists*

Each artist is a written profile that the whole studio reads: the artist card, the sitting lines, the prompt to the image model, the compiler's routing rules and the journal the artist keeps afterwards. The profile fixes a city, a medium named for a real tradition, how long the artist looks, what it notices first, what it leaves alone, and one line it says about its own work. None of this is generated while you sit, and none of it is a reading of you: every sitting line is either true of every sitting or tied to something the studio measures, such as the clock or the stillness between frames. An artist never comments on a sitter's mood, age or looks. Table 1 gives the eight as they stand, in their own words; Appendix A has their instruments and routing rules. Figure 2 shows the studio's fictional demonstration sitter drawn by all eight.

**Table 1: The eight artists in residence: the medium as the artist card names it, how long each observes, and the line each artist says of its own work.**

| Artist (city) | Medium, as the card names it | Observes | In the artist's words |
|---|---|---:|---|
| Mira (San Francisco) | Graphite economy and fog-grey wash on warm white | 11 s | I keep only the lines the fog would leave. |
| Sol (New York) | Sanguine crayon reportage on warm ivory | 8 s | Every line lands once and stays. |
| Odile (Paris) | Silverpoint-fine line, rose chalk and blue-hour wash on pale grey | 10 s | A court portrait in the old sense: the face as found, in fine silver line. |
| Aya (Tokyo) | Fine sumi likeness-line and stepped ink ranks on warm cream washi, rokusho accent | 10 s | First I look for the line. Then I let the tones stand in steps. |
| Yeon (Seoul) | Sepia line, yukrimun hatch and stepped ochre ranks on pale silk-tone | 12 s | The old rule holds: one hair wrong and it is another person, so I measure until it is you. |
| Halia (Singapore) | Nanyang ink-brush line and shophouse-pastel water-colour on shell white | 9 s | One loaded brush: thick where the form turns, thin where the light takes it. |
| Kavi (Chennai) | Kaavi red-ochre line and turmeric wash on warm ivory | 9 s | Small placements first; then a flowing line goes looking for your face. |
| Ganga (Jaipur) | Kalam line and pardakht ranks, geru tint on cream wasli | 11 s | The line does all the work here; the colour only stands behind it. |

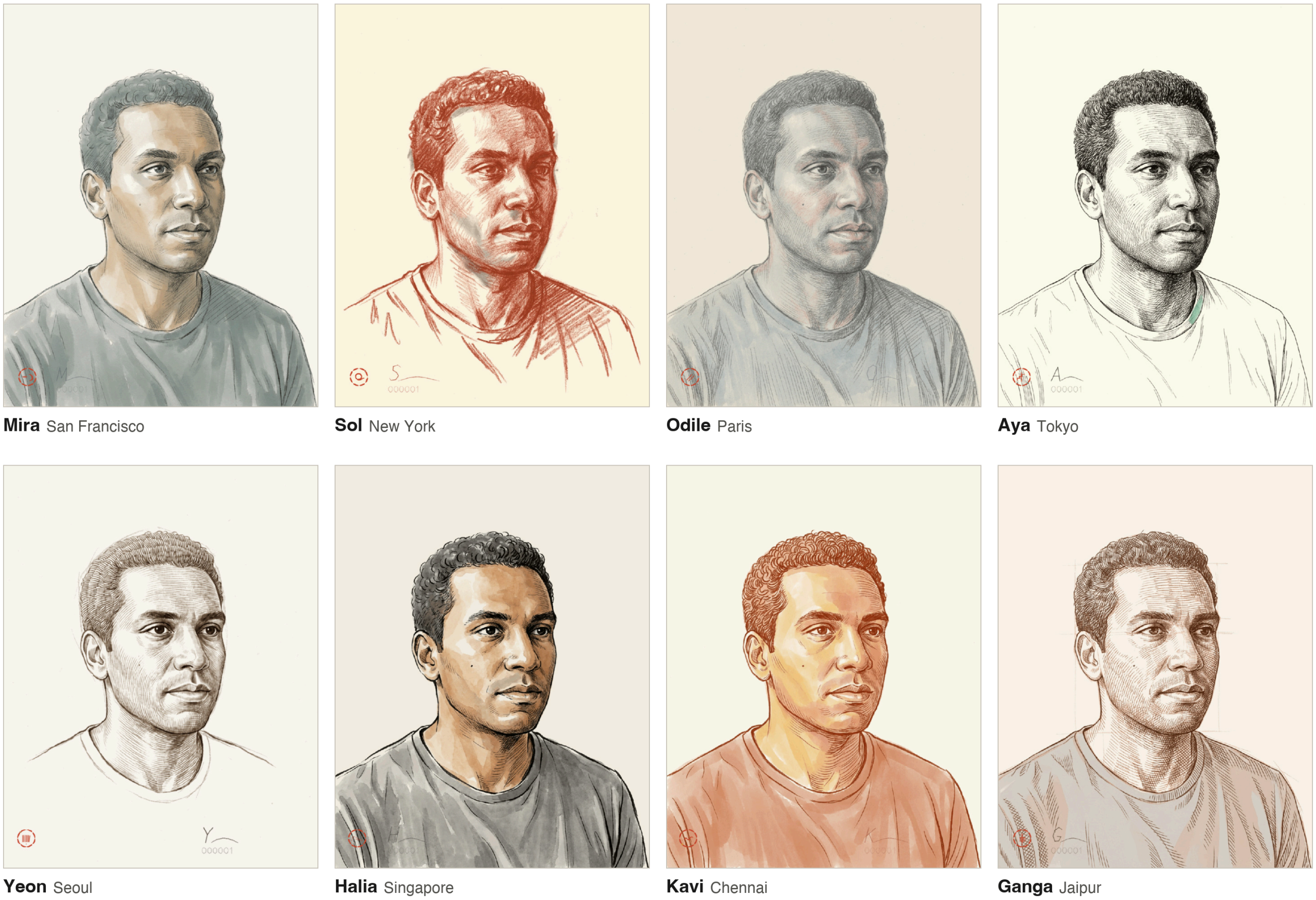

**Figure 2:** The studio's fictional demonstration sitter drawn by all eight artists, as the site's home page shows them. Each plate is the saved sheet of a drawing document compiled by the current engine from a committed reference study of that sitter, generated from one photograph under prompt versions 4 and 5; the sitter is not a real person. The signature at the foot carries the artist's initial, the edition number 000001 and the studio's sitting mark.

## 3. Related work

Stroke-based rendering has turned pictures into placed strokes since Haeberli's Paint by Numbers [10] and Hertzmann's painterly rendering with curved strokes of several sizes [12]; Hertzmann's later survey covers the family [13]. Winkenbach and Salesin [29], and Salisbury and colleagues [24], did the same for pen and ink, with prioritized stroke textures exposed as a small set of named values. Kang, Lee and Chui [15] extract coherent lines from photographs along a smoothed flow field. That literature optimises the finished picture, and the strokes are a means to it. Here the order and timing of the marks are the work itself, because a person watches them happen and later watches them again. The earlier compiler in this paper sat inside that tradition, with Zhang and Suen thinning [31], a chamfer distance transform [3], Ramer, Douglas and Peucker simplification [22, 7] and structure-tensor hatching. The current engine is closer to colour quantisation and vectorisation. It quantises with k-means [17], much as Heckbert did for frame buffers [11], and traces exact pixel boundaries, a cruder cousin of curve-fitting tracers such as Potrace [26].

Learned models produce or infer a drawing process directly. Sketch-RNN [9] learns stroke sequences from human sketches. CLIPasso [28] optimises a handful of vector strokes against a semantic objective. Learning to Paint [14] and Paint Transformer [16] predict brush strokes that reconstruct an image, one by reinforce-ment learning and one feed-forward. Zhao and colleagues [32] synthesise a plausible time lapse of how a finished painting might have been made, and call the result many possible pasts rather than the real one. Fu and colleagues [8] estimate a plausible drawing order for a finished line drawing from principles of drawing cognition and animate it; this system also authors an order for a finished picture, by fixed rules rather than inferred principles, and keeps the order as the product. My system does no learning of this kind. It authors one fixed process by rule from a finished study, and it makes no claim that the process is the model's or a person's. Because a paying sitter is told what is happening, the process has to be one I can state in full and replay exactly.

Drawing machines that perform for an audience are an older idea. Tresset and Fol Leymarie's Paul [27] draws sitters from a live camera with a physical arm and a feedback loop; Colton's Painting Fool [6] is built to be judged as a painter. Paul observes while it draws. This studio observes for a few seconds, then draws from a study that is already finished, and its artists are fictional. Berger and colleagues [1] recorded artists sketching faces under time limits down to fifteen seconds; no such recording underlies this system's order.

Skin-tone bias in imaging has a long record. Roth [23] traces how film stock and its calibration cards were balanced for light skin. Buolamwini and Gebru [4] measured commercial gender classifiers failing most on darker-skinned women, and Monk's ten-point scale [20] is now a common reference for scoring a face's tone. Yee, Tantipongpipat and Mishra [30] audited a saliency-based image cropper and found that taking the single most salient point amplifies disparities. Luccioni and colleagues [18] measured over-representation of whiteness in text-to-image models. The defect I report is on the same axis but in a different place. A per-tile adaptive threshold in the family of document-binarisation methods used by Sauvola and Pietikäinen [25] fails inside a large uniformly dark passage, because a tile's 80th percentile only reaches the paper level when about a fifth of the tile is still paper, so the estimate sits on the skin itself.

For provenance, C2PA [5] signs a manifest of how an image was made. This system signs nothing and records no provenance for the step from still to study. What it has is narrower: one document is written once and every surface reads it, so what was watched and what was kept come from the same data. For documentation, model cards [19] ask that a released model carry a plain statement of what it was evaluated on and where it fails. The nearest thing here is mechanical. Some of the sentences the product may say about itself are checked by tests (Section 7), and this paper lists where that check failed.

## 4. System

### *4.1 Design: the still, the prompt and the study*

Each sampled frame gets a score of 0.27 sharpness plus 0.22 brightness plus 0.18 composition plus 0.18 framing plus 0.15 contrast, with composition and framing from the first detected face when there is one, otherwise from visual saliency. A frame is unusable if no face is found, if brightness is below 0.22 or if sharpness is below 0.18; a second face is not a failure. Stillness, the mean absolute change between consecutive frames on a 24 by 30 luminance grid, is not a score term; it gates automatic capture alongside floors on brightness, contrast, composition and framing, and is passed to the server as one number for the title and note. After ranking, a local landmark reader checks the top four frames for a blink and ranks open-eyed frames first; if every frame blinked, the top one is kept and the sitter told.

One still leaves the device per attempt: drawn at 1,280 by 1,600 pixels, mirrored, reframed around the detected face and encoded as a JPEG at quality 0.95. With it go six summary numbers (the hour, the face count, and the framing, light, sharpness and stillness scores) and an idempotency key. On the server the still is limited to 8 MiB, 8,192 pixels a side and 40 million pixels, its declared and detected types must agree, and every metadata segment is stripped before decoding. At most 40 generations run at once, and a sitting gets at most two technical attempts.

One request goes to a hosted commercial image model per attempt, carrying the still and a fixed prompt in three parts. A foundation asks the model to redraw the sitter as a hand-drawn portrait on bare paper, with an exact likeness, keeping everyone and every animal present, removing the room, composing the head large with a modest top margin, and setting value by each face's own complexion. The artist's medium prompt names a historical register, a paper, a line, a hatching or wash method and a palette; the eight run from 669 to 2,412 characters. A traceability tail asks for distinct, confident strokes, no photographic shading and bare background paper. The prompt is at version 7. Version 7, from October 2026, restrained Odile's rose chalk to where the photographed skin actually carries warmth and set the head's top margin; it did not touch the complexion text. The request carries no seed, and there is no redraw: one design attempt per generation. The returned study is limited to 8 MiB, 2,048 pixels a side and 4 million pixels, and interlaced files are rejected. The design call has 38 seconds; the whole preparation, including queueing and the compile job, has 75. After the compile the study buffer is zero-filled. The still's buffers are zero-filled once the whole generation settles, after the compile and the sheet render, or when the deadline fires; base64 copies of it are strings and cannot be overwritten. The sitter never sees the study.

The seed is a 32-bit FNV-1a hash [21] over a sample of the study's encoded file bytes, every floor(length / 4096)th byte, from the standard offset basis with the standard prime. Nothing about the sitter enters it. The same file bytes always give the same seed; the same pixels in a differently encoded file would not.

### *4.2 Compile: the current engine*

The compiler is ordinary code. It runs once per attempt in a worker whose 25 second budget covers the compile and the sheet render, and it produces one document tagged with the engine's version. Its stages follow, in order. Three inputs come from the earlier compiler, which Section 4.3 only condenses: the candidate tone directions, the signature strokes and the rendered paper colour. Everything else here can be rebuilt from this paper.

The study is decoded and any photographed paper edge is trimmed once. A luminance drop of at least 8 levels within the outer 6 percent of the left, right or top edge marks a step. The crop moves past it by 0.4 percent of the span, rounded up, at most 8 percent per edge. The bottom inset copies the top, and the crop refuses to remove more than half of either axis. A raster larger than 1,168 pixels on a side or 912 by 1,168 pixels in area is resampled by area averaging, keeping its aspect ratio, down a ladder of maximum sides (1,168 only if already within bounds, then 960, 768, 576, 384), dropping one rung only when a complexity budget is exceeded. Of the 16 committed test studies, 14 compiled at 912 by 1,168 and two fell to 749 by 960.

A paper colour for classification is the 65th percentile, per channel, of one pixel per 8 by 8 cell, taken wherever the pixel is in the top 6 percent of rows or the outer 2.5 percent of columns on either side. Each pixel's value is the maximum of its three channels. A pixel is paper if no channel differs from the paper colour by 13 levels or more. Among the rest, a pixel is ink if it is more than 24 levels darker than the brightest pixel within 5 pixels of it, found with a separable 11 by 11 maximum, and otherwise it is tone. The two masks are disjoint and between them cover every non-paper pixel except the one-pixel image border. Because ink is a local-contrast test, the interior of a uniformly dark region wider than about 11 pixels is tone. Pixels that differ from paper by more than 40 levels extend a bounding box that only the ordering rules use; an empty box is an error, because the study contains no measurable drawing.

For each material, the in-mask pixels are quantised to a palette of at most 48 colours for ink and 80 for tone. The pixels are binned on a 33-level grid per channel (each channel divided by 8 and rounded), and each bin keeps its mean colour and count. Centres are seeded deterministically: the first is the mean of the most populated bin, and each next centre is the bin that maximises its squared distance to the nearest centre times the square root of its count. Seven iterations of count-weighted k-means follow over the bin means, in the decoded image's own RGB channels rather than a perceptual colour space, and each pixel takes its bin's cluster. The boundary of every colour band is then traced as closed loops on the pixel-corner lattice. Unit edges are emitted wherever a neighbour has a different label or the image ends, and walked into loops. Loops of fewer than four vertices are dropped, collinear vertices are removed, and each loop is written as a move followed by relative horizontal and vertical steps. Every band is an exact set of staircase loops, filled with the even-odd rule by every renderer. In all 16 test compiles both palettes reached their limits and every band was non-empty, so each document carried 128 bands. If the band paths exceed 6,000,000 characters in total, the compile moves down the ladder.

On the same working raster the earlier compiler (Section 4.3) is run in full and its output kept aside. Three things are taken from it: the paths of its tone strokes, as candidate directions; its signature strokes; and the document's metadata. The metadata includes the rendered paper colour, which is the earlier compiler's estimate (the average colour of a sparse sample of near-paper pixels) and not the classification colour above.

From the bounding box, the head is taken to end 76 percent of the way down the box and the eye line to sit 47 percent of the way down the head. The feature box runs from 9 percent of the head's height above the eye line to 30 percent below it, within 32 percent of the box's width either side of the centre. These are fixed fractions of the bounding box of the study's strongly non-paper pixels, the same for all eight artists. There is no face or landmark detection on the server, and the fractions only order marks; they never add any.

Ink is traced as follows. A chamfer distance transform (1 orthogonal, 1.414 diagonal) is computed inside the ink mask, and the mask is thinned with Zhang and Suen's algorithm for at most 80 iterations. The skeleton becomes an 8-connected graph in which a diagonal link is dropped when an orthogonal neighbour already connects. Unused edges are walked from every node whose degree is not two, then from any remaining cycles, continuing at each junction along the unused edge with the best direction continuity; walks under five pixels are dropped. Paths shorter than 8 pixels are discarded, every second skeleton point is kept plus the last, and each kept point gets a radius of 1.2 times its distance value, clamped to 0.7 to 20. Fragments are then joined, longest first, from both ends. A candidate endpoint qualifies when it lies within 9 pixels, the radii differ by at most a factor of 2.5, the tangents align at 0.88 or better, and the direction to the candidate aligns at 0.72 or better for gaps over one pixel, with at least 40 percent of the bridge as ink for gaps over three pixels. Joining continues while the running length is under 240 pixels, and a join is allowed when the combined length is at most 300, so a joined gesture can reach about 309 pixels. The best candidate minimises the gap plus 12 times one minus the alignment. Joining concatenates the point lists and invents no bridging points. Each ink contact stores rounded points, radii clamped to 0.7 to 30, the study's colour at its mid-dle vertex, unused by renderers, and the artist's line instrument, and it is rasterised into a binary coverage mask at pixel centres.

Traced ink is routed by rule. Let p be a path's middle point and len its length. A path is near the edge when p is more than 22 percent of the box width from the centre line or more than 16 percent of the head's height above the eye line. By default a path goes to pass 0 if p lies above the head's end, the path is near the edge and len exceeds 45 pixels; otherwise to pass 1 if p is in the feature box; otherwise to pass 2. Four artists change this, and a fifth changes only the instrument. For Sol any traced ink longer than 85 pixels opens in pass 0. For Odile no traced ink opens: the default openers move to pass 1 and pass 0 is left to grey tone. For Aya pass 0 is feature-box ink above a line 9 percent of the head's height below the eye line, and the default openers move to pass 1. For Kavi pass 0 is feature-box ink shorter than 22 pixels, and the default openers move to pass 1. For Yeon pass-0 ink is drawn with charcoal and the rest with her fine line; her routing is otherwise the default. Within a pass the rank is one combined score, so length only breaks ties: top-down in pass 0; the eye band left to right, then the lower features, in pass 1; and raster order over 100 by 140 pixel cells in pass 2.

Tone contacts borrow their directions from the earlier compiler's tone, wash, ground and hatch strokes of at least two points: the middle point of each is tested, and the stroke skipped unless that point is tone and not yet covered. Its path is copied unchanged, each point gets a radius of 1.5 times the old width, at least 2 and at most 12 for a sable tool or 5 otherwise, and it is drawn with the artist's tone tool. It goes to pass 4, "The colour already in the study", when the chroma at its middle point (the largest channel minus the smallest) exceeds 32; to pass 3 otherwise; and, for Odile only, to pass 0 when the chroma is below 25. A borrowed path marks its pixels covered, so a later stroke whose middle point is already covered is skipped.

Small returns close what is left. Each material is scanned on a grid from the top-left corner, in steps of 3 working-raster pixels for ink and 9 for tone. Wherever a cell holds any mask pixel that is still uncovered, one two-point diagonal contact is added from the cell's corner to the opposite corner, with a radius of 2.2 pixels for ink and 6.8 for tone. At those radii the capsule covers its whole cell. Ink returns go to pass 1 when their corner is in the feature box and to pass 2 otherwise; in pass 1 their rank scale puts them after all of the traced ink, and interleaving with traced ink by rank happens only in pass 2. Tone returns all go to pass 5, "Small returns". Across the 16 test studies these grid returns are 73.5 percent of all contacts (Section 5).

Rows are sorted by pass, then rank, then longer path first, with a stable sort. The earlier compiler's signature strokes are appended as pass 6: the artist's script initial, a trailing flick whose slant, scale and length vary with the seed, the edition number as digit glyphs under the initial, and the studio's sitting-token mark in vermilion near the foot. The signature goes to the left foot only when that foot holds less than four fifths of the ink the right one holds, measured on the earlier compiler's strokes, and to the right foot otherwise. After sorting, paths are rounded to integers, widths to 0.05 and opacities to 1/255.

Timing is arithmetic. Each artist has a rhythm row of seven pass durations summing to 35,000 ms, scaled proportionally to the 44,000 ms clock with the cumulative boundaries rounded so that the total is exact (Appendix A, drawn in Figure 3). Within a pass with budget B, an opening of 350 ms (pass 0) or 80 ms and a hold of 600 ms (pass 6) or 120 ms are set aside, leaving A. Each contact is weighted by the square root of its path length plus one. The first two contacts of passes 0, 1 and 2 get a floor of 180 ms and every other contact 0.3 ms. A contact's duration is its floor plus its weighted share of 0.88 A minus the sum of the floors, and every lift between contacts in the pass is 0.12 A divided by the number of contacts. The first contact starts at the pass start plus the opening, and each next one starts when the previous ends plus the lift. The last contact ends about 43,391 ms in; the trailing lift and the 600 ms signature hold fill the rest of the 44,000 ms total. An empty pass keeps its window and draws nothing. The live player eases each contact along its path with a cubic ease and lifts the drawn instrument between contacts.

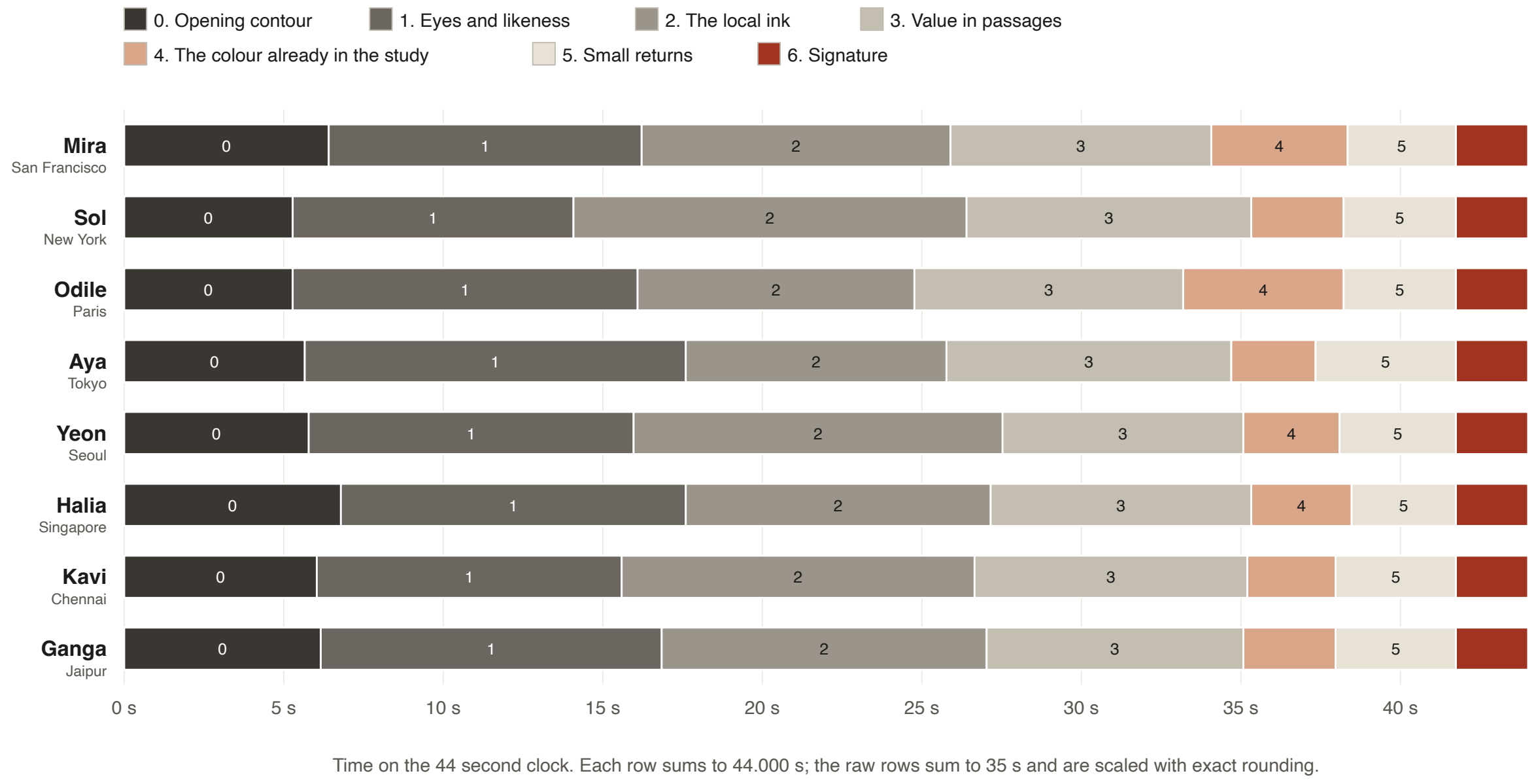

**Figure 3:** The rhythm of the eight hands: each artist's seven pass windows laid along the 44 second clock, from Table A3. The windows are fixed per artist and do not depend on the study; what changes between sittings is how many contacts each window holds. The legend gives the common pass names, and Table A2 has the names each artist actually shows.

A compile exceeds its budget, and drops one rung, when the band paths pass 6,000,000 characters, the strokes 60,000, the points 500,000, any path 65,535 points, or the packed document 8 MiB less 64 KiB. Every reader validates the document on load: one to eight materials, at most 512 bands, well-formed band colours and paths, coordinates within 4,096, a mark for every stroke, radii within 0.1 to 80, each contact starting no earlier than the previous one's end plus its lift, durations of at least 0.1 ms, each contact inside its pass window and the last end within the total. The stroke rows are packed into ten-byte columns with delta-coded paths in base64; the materials and the timed marks travel as plain JSON, and in the measured case they were 91 percent of the packed bytes.

### *4.3 The earlier compiler, condensed*

The earlier compiler still runs inside every compile, and the complexion history (Section 8.2) needs its outline. It estimated the paper level per 48 pixel tile as a high percentile of the tile's luminance, interpolated the estimates bilinearly, and clamped each so that it could not sit more than 26 levels below the sheet's global paper level. That clamp is the fairness fix of Section 8.2. A pixel's owed ink was its departure from the clamped paper level divided by the absolute range from paper to true black, never by the darkest pixel present. Line work was the strongest ridges against their neighbourhood mean, up to a per-artist share of the sheet (Appendix B1); the rest was closed morphologically and split by density against a per-artist gate into fused tone. Lines were thinned with Zhang and Suen, given width from a distance transform and simplified with Ramer, Douglas and Peucker. Tone was rebuilt as hatching along the structure tensor's dominant orientation, at per-artist thresholds and temperaments, with washes and a tonal ground for some hands (Appendix B1, B2). An ink-budget loop moved rank spacing, never opacity, toward a per-artist coverage target tilted per study by the face's own study value within a 15 percent band. Strokes were sliced into seven passes with seeded routing, and signed.

In the current engine none of its marks is drawn except its signature. Its tone, wash, ground and hatch paths are the candidate directions of Section 4.2, its signature is copied, and its paper colour and metadata are carried. Its opacity rules and its ink colour for marks no longer reach the sheet. The parameters that still shape which paths are borrowed are listed in Appendix B.

### *4.4 Perform: four surfaces, one document*

Every renderer draws a non-signature contact the same way: its shape is the union of tapered capsules along its path, one per segment, with the recorded radii, filled with a pre-rendered image of its material's colour bands rather than the contact's own colour or width. On the canvas the fill is a pattern built from an offscreen sheet on which every band of that material is painted in its palette colour. In the SVG each material is a group of band paths, and for every pass and material a clip path holding all of that pass's contacts is used to draw the group through. Because ink and tone pixels are disjoint and a material is transparent outside its own pixels, a contact shows the quantised study colour inside its capsule and nothing outside the material. The recorded colour, width, opacity and kind of a material contact are unused by these renderers; the instrument is used only to draw the tool. Signature strokes are drawn as ordinary stroked paths with per-instrument width and alpha and a seeded jitter.

**Table 2: The surfaces that read a drawing document, and their clocks.**

| Surface | How it draws | Clock |
|---|---|---|
| Live sitting canvas | Completed contacts are committed to an offscreen sheet in order; the active contact is drawn partially along its arc each frame, with a drawn instrument that lifts between contacts; flat paper | 44 s; a stage controller advances by each pass's duration |
| Replay on the same page | The same component | 0.5, 1 or 2 times: about 88, 44 or 22 s; the first performance is always at 1 |
| Shared public page | The same component, after in-browser retiming to 44 s | 44 s |
| Homepage demonstration | The same component on a fixed sample document, not retimed | 35 s (a known gap, Section 7) |
| Saved sheet (server PNG) | The SVG export rasterised natively at the document's own size | none |
| Vector download (SVG) | Generated in the browser from the retimed document: band groups, a clip path per pass and material, per-contact timing attributes, the signature last, and a description calling it an authored drawing sequence; it does not animate itself | timing as metadata only |
| Film | The browser's media recording API captures a 1,080 by 1,920 canvas at 30 frames per second (MP4 where supported, otherwise WebM), using the same partial-contact drawing code as the live canvas | 48 s: 0.6 s of blank paper, the 44 s drawing, a 3.4 s hold |

Renderers can still differ; Section 8.4 has the scope of what has and has not been checked between them. Canvas and film rasterise the band pattern at one pixel per document pixel, scaled by the device transform, with each contact composited and anti-aliased separately; the PNG and SVG render the polygons as vectors, with a pass's contacts unioned into one clip path and anti-aliased once. The canvas paints strictly in order; the SVG groups by pass and material and paints the signature last, which can matter only at anti-aliased edges and where the signature overlaps drawing. Whether every material pixel ends fully painted after anti-aliasing is untested. All surfaces paint flat paper in the document's paper colour.

Contact documents compiled in the engine's first hours in production carry 35 second stage durations, and each pass is rescaled to the 44 second rhythm when such a document loads: geometry, colours, order and materials are untouched and the stored bytes are not rewritten. Portraits made before the contact engine are earlier-compiler documents with their own 35.4 to 45 second rows; they keep their clock, except on a shared page, which plays them at half their stage time. Retiming is not the same as recompiling. On the homepage sample, a fresh compile reproduced all 11,967 non-signature contacts exactly, and the retimed stage durations. Per-contact starts differed by up to 176 ms and durations by up to 46 ms, because proportional scaling stretches the openings, holds and floors that a fresh compile keeps fixed.

## 5. From drawing marks to reconstructing pigment

The earlier compiler produced drawings in the artist's medium from the study's darkness. They were lighter and coarser than the studies. In the re-measurement of Section 8.3, its sheets carried 35 to 77 percent of the study's face darkness for the deep-complexioned sitter and 33 to 85 percent for the fair one. In the first week of September 2026 I reviewed a prototype that preserved the study and found that keeping the study's pigment kept its strongest details. I rebuilt the engine on that basis and put it into production the same month. The current engine's sheets carry 94 to 97 percent (deep) and 80 to 95 percent (fair) of the study's face darkness as the standing test measures it, against the raster's own 90th percentile; with no paper reference the figures are 99 to 100 and 98 to 100 percent, so the shortfall is mostly the sheet's flat paper, which for Odile's fair studies sits about 15 levels below the study's 90th percentile.

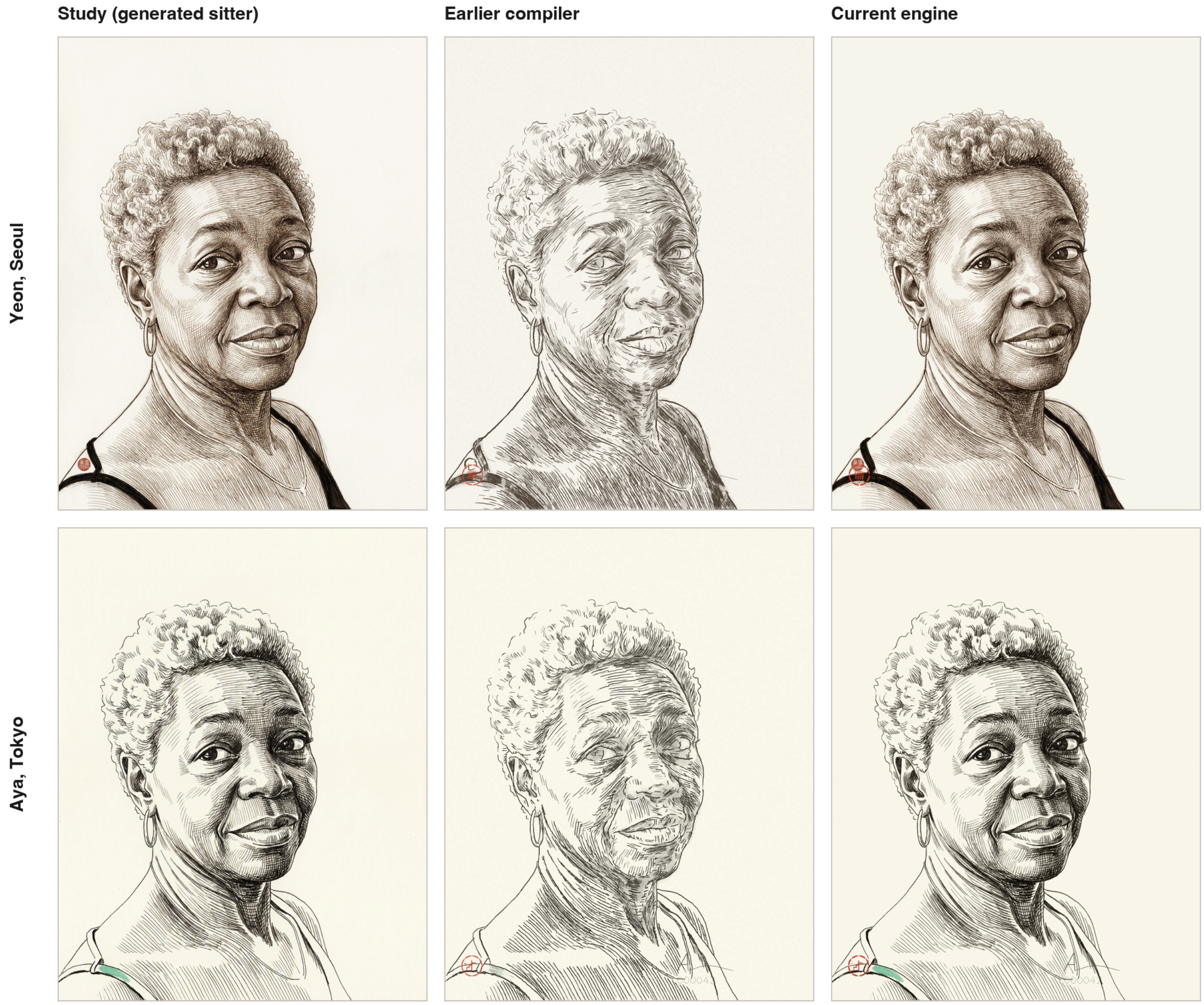

**Figure 4:** Two compiler generations on one study. Rows: Yeon (Seoul) and Aya (Tokyo), each from the committed deep-complexion generated study number 0 of the 80-study corpus, an older-looking generated sitter with grey hair and jewellery. Columns: the study, shown here only because the sitter is generated; the earlier compiler's sheet; the current engine's sheet, compiled from that same study. In Aya's row, the strap colour from the study survives on the current engine's sheet, drawn in pass 4, and is lost on the earlier compiler's. Complexion separation figures elsewhere in this paper compare this generated sitter against a separately generated fair-complexion sitter, a younger person with different hair and dress, not a single sitter with only skin tone changed.

Apart from the signature and the flat paper, every colour and value on the sheet is now a study pixel, quantised to at most 128 colours and traced as a polygon. The compiler adds an order and a clock. It cannot correct an invented accent, a changed garment colour or a poor likeness in the study. It has no rule that lightens or darkens a face: its only value changes are the 13-level paper threshold, which flattens near-paper pixels into one estimated paper colour, and quantisation to at most 128 cluster means. What the compiler still owns is the performance: which regions appear first, how the head is placed for ordering, how long each pass takes, and which instrument is drawn.

On the 16 committed studies used here, draw 0 of each hand's fair and deep sitter in the same 80-study corpus, a sheet holds 8,784 to 16,378 contacts, 195,865 in all. Pooled, 49.7 percent are 3 pixel ink returns and 23.8 percent are 9 pixel tone returns. So about three quarters of what the sitter watches is a fixed grid of two-point dabs, closing the cells that the tracer and the borrowed paths left uncovered. Traced ink lines are 20.7 percent, paths borrowed from the earlier compiler 5.6 percent, and the signature 0.2 percent. The median contact lasts 0.97 to 2.17 ms, and 96.8 to 99.7 percent of contacts (99.1 percent pooled) are shorter than one frame at 30 frames per second. At 20 seconds, 31 to 52 percent of a sheet's contacts are complete; at 35 seconds, 61 to 82 percent. Sol's rhythm leaves a pause. His pass 3, "Dry crayon weight", was empty on both of his test studies, because tone returns always go to pass 5 and every borrowed path in those studies had enough chroma to go to pass 4. With no contact active from the end of pass 2 through the start of pass 4, the canvas held still for 9,126 ms: pass 2's 120 ms closing hold, the 8,926 ms of empty pass 3, and pass 4's 80 ms opening. Odile and Aya have similarly stretched passes.

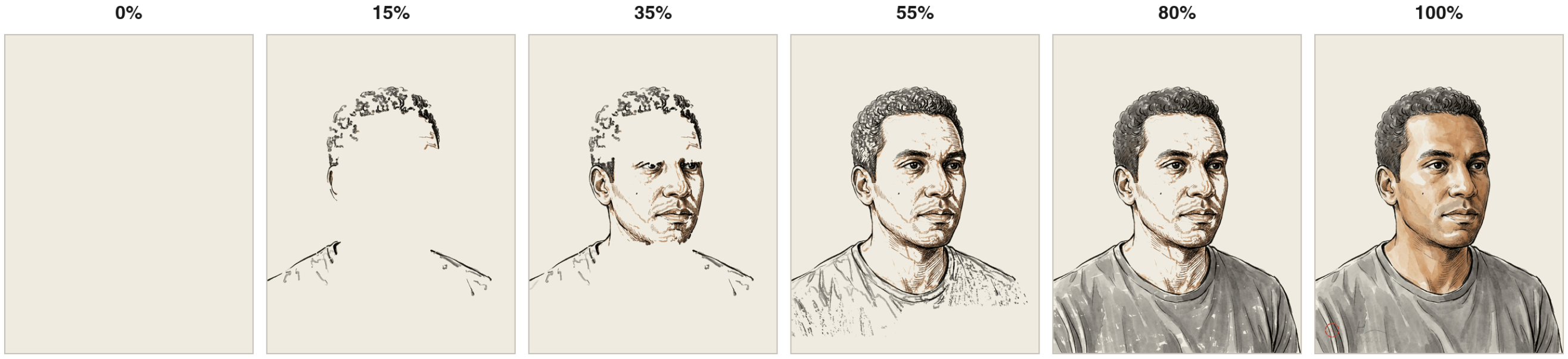

**Figure 5:** Six frames of one Halia (Singapore) drawing at 0, 15, 35, 55, 80 and 100 percent of its 44 second clock, rendered offline from the drawing document itself (10,198 contacts). The sitter is the studio's fictional demonstration sitter, compiled from the committed Singapore reference study with the production seed. Each frame is an exact state of the document, not a screen recording.

So the performance is an authored reconstruction at a compressed speed. It is not a recording of a hand, not the image model's own process, and not an artist choosing anything while it plays. The vector export says the first two of those in its own description element, and the site's account of what happens when you sit says the same. The home page's one-line summary contradicted it for two months, and Section 7 records that.

## 6. Determinism and replay

The compiler contains no wall clock and no unseeded randomness. The only pseudo-random generator is a small seeded one inside the earlier compiler, used for hatch angles and region routing; the signature's slant, scale and flick take their variation from the seed. The output is a pure function of the study's decoded pixels, the seed (from the encoded file bytes), the artist's configuration and profile, and the edition number, which is drawn into the signature. Iteration over the compiler's maps and sets follows a raster scan or a fixed loop, and the sort is stable.

I compiled one committed study (Halia, draw 0, with the production seed and edition number 000001) in four separate processes on one desktop machine: three on one runtime release and one on the next minor release, all on the same JavaScript engine build (V8 13.6). The packed document's SHA-256 was identical in all four: 8,784 contacts, 43,830 path points, 128 bands, 4,197,804 bytes packed and 882,593 bytes gzipped. Changing only the seed to 42 gave 8,818 contacts and a different hash. Passes 0 to 2 were untouched; the borrowed paths in passes 3 and 4 changed, the tone returns in pass 5 followed their coverage, and the signa-ture changed because its slant, scale and flick are seeded. Changing only the edition number to 00123 removed one signature stroke and changed the hash. The committed test suite asserts deep equality of two compiles in one process.

The scope is byte-identical recompilation on one machine and one engine build, from identical file bytes, configuration and edition number. It is not verified across machines or engine builds, and in particular not for the production container, which runs a different operating system and a different runtime major version. The language standard leaves the maths the compiler uses (hypot, pow, atan2, sine and cosine) implementation-approximated, so identity across engine builds is neither promised by the standard nor tested here. Determinism also holds only from the study onward. The model call carries no seed, and the same still can return a different study.

Replay makes no model call and no compile. The stored document is served, gzipped when the client accepts it, and the same player component runs it. The document is not small: 4.20 to 8.11 MB of JSON on the 16 test studies, 0.8 to 1.6 MB compressed, and the largest was 97.4 percent of the compile budget. Replay can change the clock, by speed and by the retiming of older documents, and never the geometry or the order.

## 7. Claims pinned by tests

I write the sentence the product is allowed to say as a test against the product, so that a change to the product without a change to the sentence fails the build. Tests read the user-facing copy and fail on claims about a sitter's mood, age or character. A test checks that a number shown as a count is computed as a count and not from a high-water mark. For the current engine there are tests that a recompile in one process is equal, that packing preserves the materials and every path, that the seven stage durations sum to exactly 44,000 ms for all eight artists, and that a packed document stays under 8 MiB. Others check that the film plan equals the contact starts, and that the SVG export carries the timing attributes and the tool, embeds no raster and no external reference, and contains the sentence calling it an authored drawing sequence. Others check that the reference manifest declares eight distinct studies, one compiler, no geometry edits and fictional sitters, which is a check on the declaration rather than on the plates; that retiming an older document does not mutate it and is idempotent; that the preparation deadline covers the design and compile budgets; and that the container build emits a document whose marks equal its strokes and a PNG of the document's size. A browser test asserts that choosing double speed on a replay sends no second generation request.

It has failed in six places. For some weeks the site's explanation said the film was "about 39 seconds"; that was true at the 35 second clock, the code now makes 48, and no test read that sentence. It left the site in a copy rewrite, not because a test caught it. The standing complexion test compiles every study with the earlier compiler, so its floors guard a compiler that production no longer draws with, and the re-measurement in Section 8.3 is a one-off run rather than a test. An earlier version of this paper said its parameter table was generated from the source; the generator read only the manuscript, and three of the table's rows had drifted from the code by the time the engine changed. The homepage demonstration still plays its sample at 35 seconds while every sitting plays 44, and its button says so. A fifth failure belongs here too: the withdrawn "roughly half the rate" claim of Section 8.2 stayed published on the site's own paper page, untested, from August 2026 until the site revision that accompanies this paper. A sixth: the home page summarised the sit-ting as "stroke by stroke, down to the signature" for two months after Section 5 was true, and that line is corrected in the same site revision.

The pass names have gone wrong the same way twice. The earlier compiler once had passes named for the face, and the names implied that the compiler routed marks from the measured face into them. It did not, the names came down, and some weeks later a head-finding rule was built and the names went back. The current engine routes with fixed fractions of the bounding box of the study's strongly non-paper pixels for all eight artists, so a pass named "Eyes and likeness" holds the marks whose middle point, or whose cell corner, falls in a band of that box, plus whatever each hand's routing rule moves there, not marks from detected eyes. Sol's pass named "Dry crayon weight" was empty on both of his test studies. The sitter reads these names while the pass plays. Table A2 lists them as they are.

## 8. Evaluation

### *8.1 What is measured*

Determinism is reported in Section 6, with its scope. Document statistics come from compiling the same 16 studies as Section 5, on one desktop machine with the production seed formula.

**Table 3: Per-artist document statistics on the 16 committed studies (draw 0 of the 80-study corpus). "Under one frame" is the share of contacts shorter than 33.3 ms; MB is packed JSON before compression, at 10^6 bytes.**

| Artist | Contacts, fair | Contacts, deep | Under one frame, fair | Under one frame, deep | MB, fair | MB, deep |
|---|---|---|---|---|---|---|
| Mira | 10,976 | 10,053 | 99.0% | 98.7% | 5.07 | 4.74 |
| Sol | 11,455 | 16,378 | 99.1% | 99.6% | 5.38 | 8.11 |
| Odile | 12,915 | 14,394 | 99.7% | 99.7% | 5.92 | 7.57 |
| Aya | 12,269 | 11,032 | 99.0% | 98.5% | 4.36 | 5.12 |
| Yeon | 16,049 | 15,764 | 99.6% | 99.5% | 6.50 | 7.82 |
| Halia | 9,320 | 8,784 | 98.8% | 96.8% | 4.48 | 4.20 |
| Kavi | 11,009 | 9,156 | 99.7% | 98.9% | 5.20 | 5.01 |
| Ganga | 13,966 | 12,345 | 99.1% | 99.1% | 5.86 | 5.31 |

Compile took 2.3 to 6.0 seconds per study on that machine (median 3.25 s); in the fairness run the median was 3.10 s, against 1.75 s for the earlier compiler. These are desktop numbers, not production latency. The only production samples are one live generation at 32.6 s including retrieval and one container canary at 11.2 s including the PNG.

### *8.2 The complexion history on the earlier compiler*

On 2 August 2026 I ran six generated sitters through the real design prompts (14 studies and 5 re-runs). Two line-first hands left a deep-complexioned sitter's face as untouched paper, and one watercolour hand under-saturated deep skin by one to two value steps. I changed the prompts, adding a register-shift clause for four hands and a complexion clause to the shared foundation text, and closed the report on an assumption I never checked: that because the compiler samples darkness from the study, a fix upstream would reach the sheet on its own. That study-level effect has not been separately re-measured since (Section 8.3 has the corpus's own prompt-version gap).

On 15 August a second audit found the defect in the compiler. Ink was measured against the 80th percentile of each 48 pixel tile, and on a uniformly dark face that percentile is the skin itself, so much of the face was measured as paper. Measured as the share of the study's face value surviving the trace, deep against fair, the eight hands gave Mira 38 against 51 percent, Sol 33 against 46, Odile 13 against 43, Aya 32 against 42, Yeon 28 against 42, Halia 48 against 56, Kavi 44 against 55 and Ganga 34 against 41. The deep sitter was lower in all eight, the mean gap was about 13 points, and five of the eight sheets drew the deep sitter lighter than the fair one. An earlier version of this paper, published on the studio's site until this revision, says the deep face survived "at roughly half the rate", and the same sentence is still in the compiler's comments and the standing test's header. The ratios are 0.30 to 0.86 with a mean of 0.71; only Odile falls to half or below, and I withdraw "roughly half" as a general statement. There were three fixes: clamp the local estimate so it could not sit more than 26 levels below the sheet's global paper level, normalise owed ink against the absolute range to true black, and make the ink budget spend rank spacing rather than opacity. After the clamp, the mean gap fell to about 5 points with mixed direction. Whether each fix was sufficient alone was never tested, so I do not claim it. A standing test followed: two generated sitters, all eight hands, deep darker than fair everywhere, a separation of at least 1.0 per hand and a roster mean of at least 4.0. Loosening the clamp to 999 levels made all three assertions fail. Per-hand floors were set at three quarters of each hand's value on frozen 456 by 584 fixtures measured around 16 August (the same eight numbers stand as Table 4's Floor column).

On 23 August I replaced the frozen fixtures with 80 fresh studies at production size, five draws per sitter per hand under prompt version 4. The frozen test was green while three hands were under their floors on fresh work: San Francisco 5.5, New York 3.9, Paris 6.1. The root cause was a single ink target per hand for both sitters, which pulled the two complexions toward one coverage. A per-study tilt of the target within a 15 percent band fixed it: those three became 7.6, 5.7 and 8.9, and the roster mean went from 8.96 to 11.24. The same day I withdrew a claim that most of the complexion information was "lost in the trace" (a roster retention of about 40 percent). It had divided a sheet measurement by a study measurement taken in a different pixel frame and a different medium; normalised by each side's own mean, the retention was 83 percent. What was left was one hand (Ganga) and one mechanism (a global line quantile starving a deep face in the three fine-ranked hands, smallest line share, Table B1). The fix I tried cut Ganga's traced line two- to threefold, so I rejected it. The earlier compiler was revised again on 2 September 2026, so those 24 August values are not the same compiler as the earlier-compiler column of Table 4. Its test still passes at the current source, five of five on 17 September 2026. All of this measured the earlier compiler.

### *8.3 Re-measurement on both engines*

On 17 September 2026 I compiled all 80 corpus studies through both engines and rendered their sheets with the production PNG path, offline, with no model call. The measure is the standing test's own: the mean over a face box of how far each pixel's luma (gray = 0.2126R + 0.7152G + 0.0722B) sits below the raster's 90th-percentile luma, taken as the paper. The separation for a hand is the median over the five deep draws minus the median over the five fair draws. The two sitters differ in age, sex, hair and dress as well as complexion, so every separation number below measures the difference between two different generated people, of whom one is darker, and not an isolated skin-tone comparison; it is a compiler-fidelity measure under that difference, not a complexion measure on its own. The study is measured in each engine's own pixel frame, after the same edge trim and, for the current engine, the same resampling, so that the retention figures di-vide like by like. Raw retention is sheet separation over study separation; normalised retention divides each side by its own mean darkness first, which removes the effect of a sheet being lighter overall. The measurement script reproduces the committed test's numbers for the earlier compiler exactly, and the current engine was run with both the test's fixed seed and the production seed formula; the two agree to within 0.01 luma levels on every hand in the test box and the central box, the regions where both were measured.

**Table 4: Complexion separation on 80 generated studies (median deep minus median fair face darkness, luma levels), and retention of the study's separation. "Current, deep kept" and "Current, fair kept" are each sitter's own share of their study darkness surviving on the current engine's sheet; the earlier compiler's own range is in Section 5. Study is measured in the current engine's pixel frame; for Sol the earlier compiler's own frame gives 8.8, the denominator of its 70 percent retention, and every retention figure divides unrounded medians before rounding for display. The floor is the standing test's per-hand minimum.**

| Artist | Study | Earlier sheet | Current sheet | Earlier retention, raw / normalised | Current retention, raw / normalised | Current, deep kept | Current, fair kept | Floor |
|---|---|---|---|---|---|---|---|---|
| Mira | 19.1 | 12.2 | 19.0 | 64% / 131% | 100% / 106% | 95% | 92% | 6.5 |
| Sol | 9.1 | 6.2 | 8.2 | 70% / 96% | 89% / 95% | 95% | 94% | 5.2 |
| Odile | 25.5 | 10.2 | 30.5 | 40% / 122% | 120% / 134% | 95% | 80% | 6.5 |
| Aya | 21.5 | 8.0 | 21.0 | 37% / 51% | 98% / 106% | 94% | 91% | 7.5 |
| Yeon | 32.8 | 6.9 | 31.0 | 21% / 33% | 95% / 100% | 96% | 95% | 1.2 |
| Halia | 32.0 | 25.1 | 31.9 | 79% / 98% | 100% / 103% | 97% | 95% | 9.9 |
| Kavi | 39.2 | 23.0 | 39.7 | 59% / 91% | 101% / 106% | 96% | 92% | 3.3 |
| Ganga | 14.6 | 5.7 | 14.9 | 39% / 52% | 102% / 107% | 96% | 93% | 3.0 |
| Roster mean | 24.2 | 12.1 | 24.5 | | | | | |

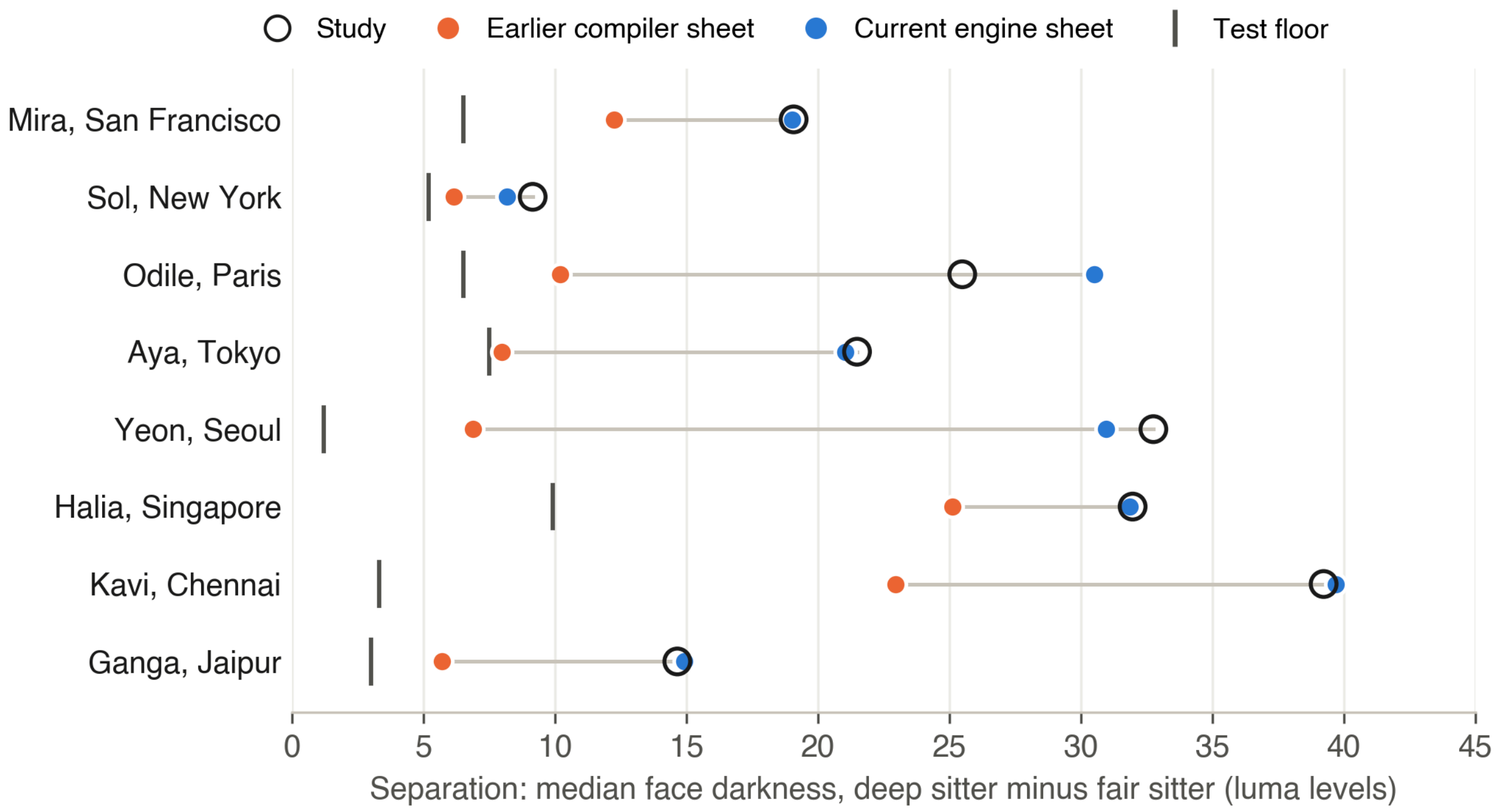

**Figure 6:** Complexion separation per hand (median face darkness of the deep sitter minus the fair sitter, in luma levels) for the study, the earlier compiler's sheet and the current engine's sheet, on 80 generated studies, with each hand's test floor marked. The deep and fair sitters are two different generated people, one older with grey hair, one younger, not one sitter with only skin tone varied; each bar is the gap between them, not an isolated complexion reading.

Both engines pass every assertion of the standing test on this corpus. The current engine separates more than the earlier compiler for all eight hands and carries about twice its separation: roster means 24.2 for the studies, 12.1 for the earlier compiler and 24.5 for the current engine. Its raw retention is 89 to 120 percent with a median near 100, and normalised 95 to 134. Odile's figure above 100 is a paper-reference effect; with no paper reference it is 99 percent. The earlier compiler kept 21 to 79 percent raw; normalised, it kept 91 to 131 percent for five hands and 33, 51 and 52 percent for Yeon, Aya and Ganga, the three fine-ranked hands. Pairing draw i deep with draw i fair, the earlier compiler inverted 3 of 40 pairs and the current engine none. Across all 25 combinations per hand, the earlier compiler inverted 23 of 200, the current engine 7, and the studies themselves 10; with no paper reference both the studies and the current engine invert 5 of 200, so that difference is a paper-reference effect, not a correction. The current engine separates more throughout; that direction holds across all four face regions (the test's box, a tighter central box, the engine's own feature rectangle from the study, and the same rectangle from each sheet's own ink) crossed with three paper references, twelve combinations in all. The magnitudes move, and two of those moves matter: with no paper reference the earlier compiler inverts Paris and Jaipur, and the current engine's retention for Ganga falls to 78 percent in the feature rectangle taken from each sheet's own ink. The only negative per-hand separations anywhere are the earlier compiler's Paris and Jaipur with no paper reference, where paper tone differs between draws.

On these 80 generated studies the current engine does not reduce the complexion difference that exists in the study. That is what its design predicts, since it copies the study's pigment rather than re-deriving value from a threshold. Under the test's paper reference it keeps 94 to 97 percent of the deep sitter's face darkness and 80 to 95 percent of the fair sitter's (Table 4); with none, it disappears, 98.7 to 99.8 against 97.5 to 100.2, fair kept marginally better in five of eight. The low end is Odile: her fair sheets carry a paper colour 14.6 levels below the study's 90th percentile, giving 79.7 percent; the other seven hands span 91.0 to 95.2 against 93.9 to 97.3. The likely cause is the reference, not the paper classification. The earlier compiler, even after the August fixes, drew the two sitters closer together than their studies were, mostly because its sheets were lighter overall, and for three hands in relative terms too. Study separation varies fourfold between hands (Sol 9.1, Kavi 39.2) before any compiler runs, and the current engine neither narrows that nor corrects anything the study gets wrong. On this one statistic, mean face darkness, the compiler no longer narrows the difference the study carries, so that part of the fairness question has moved upstream, to the prompt and the hosted model. What the compiler still owns, and what is unmeasured, is which marks are traced and when each region appears.

The sitters are two generated faces, and there is no consented real-face corpus and no Monk scale scoring. There are two complexions and five draws each, and paired separations for one hand range over tens of points. The measure is one family, mean darkness in a face region; it says nothing about likeness, features, eyes, hair or whether a face reads as the right person. The corpus was generated under prompt version 4 and production uses version 7. And the committed test still guards only the earlier compiler; nothing guards this result yet.

### *8.4 What is not measured*

No perceptual study exists: nobody has been asked whether a sheet reads as drawn, which artist made it, or whether two artists can be told apart. Likeness is not measured at all. No real faces have been evaluated. No device has been benchmarked: replay of up to 8 MB of JSON with per-frame pattern fills, and film recording, on a phone are untested. Production latency exists only as single samples. And there is no pixel-level comparison of the canvas, the PNG, the SVG and the film for a document of the current kind; every parity number on record measured a different renderer.

## 9. Limitations and ethics

Every fairness number in Section 8.3 comes from two synthetic faces, generated by that same hosted image model, and the August 2026 study-step audit (Section 8.2) from six. So "deep" and "fair" here are a model's rendering of those complexions, from a class of models with documented bias toward lighter skin [18, 2]. The two sitters were not matched for age, sex, hair or accessories: the deep one reads as an older woman with short grey hair and jewellery, the fair one as a young man with blond hair, and neither is scored on the Monk scale. This measures the compiler on the studies it is given. It does not measure how the studio draws people. Closing that gap needs consented photographs scored on a published scale, and it is the most valuable thing missing from this paper.

I have given the complexion defect more room in this paper than the engine that replaced it, and that is deliberate. It was a bug that produced no error, that a test written in good faith did not catch, and that fell hardest on the people a portrait studio should be most careful with. A studio that invites everyone to sit owes them the record of where it went wrong, and a reader who builds anything like this will meet the same mechanism in any adaptive threshold they reach for.

The study comes from a commercial hosted image model that I do not name in this paper; it is chosen at runtime from stored settings and can be swapped. It is not seeded, its studies are not reproducible, and its behaviour can change without notice. Determinism starts after it.

Fidelity now depends on the study. The sheet reproduces the study's pixels, so an invented accent, a changed garment colour or a poor likeness in the study reaches the sheet unchanged. The ordering rules assume a framed face and can misplace a profile or a wider scene. The site tells the sitter this.

Nearly every contact is shorter than a video frame and three quarters are grid dabs. A sitter sees regions filling in a fixed order at a compressed speed, and the studio should not describe it as watching every stroke.

The eight artists are fictional, and their media are real. The prompts name registers from other people's traditions, some living and some historical, among them Kamakura nise-e, late-Joseon chosanghwa, Nanyang ink brush, kolam and kaavi wall drawing, and the khaka stage of a Jaipur miniature, alongside Clouet's court drawings, Ashcan crayon reportage and California Tonalism. The studio borrows their names and methods through a text prompt to a model whose training data I do not know. I cannot resolve that, so I state it, and I have tied each residency to the place its tradition comes from: half of the proceeds of paid sittings are pledged to arts initiatives connected to the eight cities.

One still leaves the device per attempt, after consent. It is stripped of metadata and bounded before decoding, and its buffers are zero-filled once the whole generation settles; it is never written to storage. On the vendor the studio runs today, the request carries a header asking the provider not to retain the payload. That is a setting the provider offers; I cannot guarantee it on the provider's behalf, and the privacy notice says so. The alternative vendor the settings allow sends no such retention control.

I built and checked the system and this paper alone, with automated tests and internal reviews and no external audit.

## 10. Conclusion

The studio keeps one document per sitting. The live drawing, the saved sheet, the vector export and the film read it, and a replay serves that file again with no model call. The current engine makes that document from the study's own pigment, revealed by eight to sixteen thousand timed contacts over 44 seconds, in an order and at a tempo that each of the eight artists was written to have. It is a reconstruction, and I have tried to describe it at the level a reader could rebuild it. Rebuilding the compiler did not close the fairness question. The earlier compiler could erase a dark face on its own, and the fix for that is measured and guarded. The current engine keeps whatever complexion the study carries, so the next measurement has to be of the studies, on real and consented faces, and it has not been done. Until it is, the honest thing I can offer is the studio itself: anyone can sit for it at https://everyfaceproject.com, watch the drawing, and keep the file it came from.

## Availability

The studio is live at https://everyfaceproject.com, where anyone can sit for a portrait, watch it drawn, replay it, download the vector file and record the film. The register is planned at 10,000 portraits, numbered in the order they are made, and closes when it is full. The site's reading room carries a plain account of what happens when you sit, and this paper. The compiler source, the 80-study corpus and the measurement script are not released with this version of the paper.

## Acknowledgements

I designed and built the studio, the two compilers and the tests, ran every measurement reported here, and wrote this paper. AI coding assistants helped me write software and the measurement scripts faster than I could alone, and AI writing tools read drafts of this paper for errors and loose sentences. Every rule described here is one I chose, every number was checked by me against the source and the dated records, and the mistakes are mine. No external funding supported this work.

## Appendix A. The eight artists in the current engine

**Table A1: Observation, instruments and routing rules per artist. The tone tool's kind and radius cap apply to borrowed tone contacts.**

| Artist (city) | Observe | Line tool | Tone tool, kind, radius cap | Signature tool | Routing beyond the default |
|---|---|---|---|---|---|
| Mira (San Francisco) | 11 s | graphite | stump, tone, 5 px | sable | none |
| Sol (New York) | 8 s | graphite | chalk, tone, 5 px | chalk | traced ink longer than 85 px opens in pass 0 |
| Odile (Paris) | 10 s | fine | sable, wash, 12 px | sable | no traced ink in pass 0; tone with chroma below 25 opens in pass 0 |
| Aya (Tokyo) | 10 s | fine | sumi, tone, 5 px | sable | pass 0 is feature-box ink above the eye line plus 9 percent of the head height; default openers move to pass 1 |
| Yeon (Seoul) | 12 s | fine, charcoal in pass 0 | sumi, tone, 5 px | sable | pass-0 ink drawn with charcoal |
| Halia (Singapore) | 9 s | sumi | sable, wash, 12 px | sable | none |
| Kavi (Chennai) | 9 s | sumi | sable, wash, 12 px | sable | pass 0 is feature-box ink shorter than 22 px; default openers move to pass 1 |
| Ganga (Jaipur) | 11 s | fine | fine, tone, 5 px | sable | none |

**Table A2: Pass names as the compiler emits them and the sitter reads them.**

| Artist | Pass 0 | Pass 1 | Pass 2 | Pass 3 | Pass 4 | Pass 5 | Pass 6 |
|---|---|---|---|---|---|---|---|
| Mira | The long graphite contour | Eyes and likeness | The local ink | Value in passages | The colour already in the study | Small returns | Signature |
| Sol | The graphite gesture | Eyes and likeness | The local ink | Dry crayon weight | Warm crayon passages | Chalk returns | Signature |
| Odile | Prepared grey passages | Silver line and likeness | The local ink | Value in passages | The colour already in the study | Small returns | Signature |
| Aya | First likeness lines | Eyes and likeness | The local ink | Value in passages | The colour already in the study | Small returns | Signature |
| Yeon | The willow contour | Eyes and likeness | The local ink | Value in passages | The colour already in the study | Small returns | Signature |
| Halia | The loaded brush contour | Eyes and likeness | The local ink | Value in passages | The colour already in the study | Small returns | Signature |
| Kavi | Small placements | Kaavi line and likeness | The local ink | Value in passages | The colour already in the study | Small returns | Signature |
| Ganga | The measure and contour | Eyes and likeness | The local ink | Value in passages | The colour already in the study | Fine return strokes | Signature |

Sol's pass 3, "Dry crayon weight", was empty on both of his committed test studies (Section 5).

**Table A3: Rhythm: pass durations at the 44 second clock, in seconds (the raw rows sum to 35 s and are scaled with exact rounding to 44,000 ms). Figure 3 draws these rows.**

| Artist | Pass 0 | Pass 1 | Pass 2 | Pass 3 | Pass 4 | Pass 5 | Pass 6 |
|---|---:|---:|---:|---:|---:|---:|---:|
| Mira | 6.411 | 9.806 | 9.680 | 8.172 | 4.274 | 3.394 | 2.263 |
| Sol | 5.280 | 8.800 | 12.320 | 8.926 | 2.891 | 3.520 | 2.263 |
| Odile | 5.280 | 10.811 | 8.675 | 8.423 | 5.028 | 3.520 | 2.263 |
| Aya | 5.657 | 11.943 | 8.171 | 8.926 | 2.640 | 4.400 | 2.263 |
| Yeon | 5.783 | 10.183 | 11.565 | 7.543 | 3.017 | 3.646 | 2.263 |
| Halia | 6.789 | 10.811 | 9.554 | 8.172 | 3.143 | 3.268 | 2.263 |
| Kavi | 6.034 | 9.555 | 11.062 | 8.549 | 2.766 | 3.771 | 2.263 |
| Ganga | 6.160 | 10.686 | 10.183 | 8.045 | 2.892 | 3.771 | 2.263 |

## Appendix B. Earlier-compiler parameters that still shape borrowed tone directions

These values are read only by the earlier compiler, which runs inside every compile. They change which tone paths exist to be borrowed, and so the order, timing and instrument path of tone contacts; they should not set colour or value on the current sheet, which is an inference from the renderers rather than a measurement. A further parameter, widthScale, scales the earlier compiler's non-wash widths and so the radius of any tone contact borrowed from that artist; across the roster it ranges from 0.85 to 1.2; the eight values are in the source, not this paper's measured record, so Table B1 omits them. The ink colour and signature seed reach the sheet directly through the signature.

**Table B1: Tone parameters. Line share in brackets is the roster default.**

| Artist | Density gate | Spacing | First cross | Second cross | Build | Line share | Face ink |
|---|---|---|---|---|---|---|---|
| Mira | 0.50 | 3.2 | 0.60 | 0.82 | cross | (0.09) | 0.35 |
| Sol | 0.42 | 3.2 | 0.42 | 0.62 | cross | (0.09) | 0.39 |
| Odile | 0.40 | 2.8 | 0.55 | 0.76 | cross | (0.09) | 0.35 |
| Aya | 0.72 | 3.0 | 0.30 | 0.52 | close | 0.03 | 0.37 |
| Yeon | 0.82 | 4.2 | 0.32 | 0.50 | close | 0.03 | 0.38 |
| Halia | 0.50 | 3.6 | 0.64 | 0.86 | cross | (0.09) | 0.31 |
| Kavi | 0.50 | 3.0 | 0.50 | 0.76 | cross | (0.09) | 0.37 |
| Ganga | 0.85 | 4.4 | 0.30 | 0.48 | close | 0.03 | 0.36 |

**Table B2: Study reference, ground, wash and signature parameters.**

| Artist | Face study reference | Underlayer | Drag | Wash, treatment | Head pass | Signature ink (RGB) | Signature seed |
|---|---|---|---|---|---|---|---|
| Mira | 0.1220 | 0.16 | none | yes, passage | 2 | 91, 97, 103 | 19 |
| Sol | 0.1880 | 0.14 | 0.125 | no | none | 182, 92, 67 | 23 |
| Odile | 0.1116 | 0.14 | none | yes, passage | 2 | 124, 130, 136 | 29 |
| Aya | 0.1522 | 0.05 | none | yes, accent | 2 | 61, 56, 49 | 11 |
| Yeon | 0.1762 | 0.05 | none | yes, accent | 2 | 75, 64, 52 | 13 |
| Halia | 0.2002 | 0.14 | none | yes, passage | 2 | 58, 56, 53 | 17 |
| Kavi | 0.1658 | 0.12 | none | yes, passage | none | 156, 74, 51 | 7 |
| Ganga | 0.1492 | 0.04 | none | yes, accent | 2 | 91, 70, 54 | 3 |